\documentclass[twocolumn,aps,showpacs]{revtex4}
\usepackage{amssymb}
\usepackage{epsf}
\usepackage{amsmath}
\usepackage{graphicx}

\def\lsim{\lower -0.3ex \hbox{$<$} \kern -0.75em \lower 0.7ex \hbox{$\sim$}}
\def\gsim{\lower -0.3ex \hbox{$>$} \kern -0.75em \lower 0.7ex \hbox{$\sim$}}
\def\Journal #1,#2,#3,#4#5#6#7{#1 {\bf #2}, #3 (#4#5#6#7)}

\newcommand{\GVec}[1]{\mbox{\boldmath$#1$}}

\def\Vec#1{{\bf #1}}
\def\GVec#1{\mbox{\boldmath $#1$}}

\begin{document}

\title{Landau level spectra and the quantum Hall effect
of multilayer graphene}
\author{Mikito Koshino$^{1}$ and Edward McCann$^{2}$}
\affiliation{
$^{1}$Department of Physics, Tohoku University, Sendai, 980-8578, Japan\\
$^{2}$Department of Physics, Lancaster University, Lancaster, LA1
4YB, UK}

\begin{abstract}
The Landau level spectra
and the quantum Hall effect of ABA-stacked multilayer graphenes
are studied in the effective mass approximation.
The low-energy effective mass Hamiltonian
may be partially diagonalized into an approximate block-diagonal form,
with each diagonal block contributing parabolic bands except,
in a multilayer with an odd number of layers,
for an additional block describing
Dirac-like bands with a linear dispersion.
We fully include the band parameters and, taking into account the symmetry
of the lattice, we analyze their affect on the block-diagonal Hamiltonian.
Next-nearest layer couplings are shown to be particularly
important in determining the
low-energy spectrum and the phase diagram of the quantum Hall
conductivity, by causing energy shifts, level anti-crossings, and
valley splitting of the low-lying Landau levels.
\end{abstract}


\maketitle

\section{Introduction}

Since the isolation of graphene flakes \cite{novo04},
the chiral nature of quasiparticles in monolayer and bilayer graphene
has been observed in a range of phenomena including the integer quantum
Hall effect \cite{novo05,zhang05,novo06},
Klein tunneling \cite{che06,kat06,huard07,young09},
weak localization \cite{morp06,mccWL,mor06,heer07,wu07,gor07,tik08,tik09},
and photoemission \cite{ohta06,zhou06,bost07,mucha08}.
Generally speaking, the effective mass models
\cite{wallace,slonweiss,
McClure_1956a,Shon_and_Ando_1998a,Zheng_and_Ando_2002a,
Gusynin_and_Sharapov_2005a,Peres_et_al_2006a,
mcc06a,guinea06,kosh_bilayer}
of monolayer and bilayer graphene
have been very successful in describing these phenomena.
The low-energy band structure of bilayer graphene is composed
of a pair of parabolic bands \cite{mcc06a,guinea06,kosh_bilayer},
and is distinct from monolayer which
has a Dirac-like linear dispersion \cite{slonweiss,McClure_1956a}.
In a magnetic field, the Landau level structures of monolayer
\cite{McClure_1956a,Shon_and_Ando_1998a,Zheng_and_Ando_2002a,
Gusynin_and_Sharapov_2005a,Peres_et_al_2006a}
and bilayer \cite{mcc06a,castro07}
differ in the degeneracy at zero-energy,
and quantum Hall plateaus appear at different filling factors
accordingly \cite{novo05,zhang05,novo06}.

\begin{figure}
\centerline{\epsfxsize=0.85\hsize \epsffile{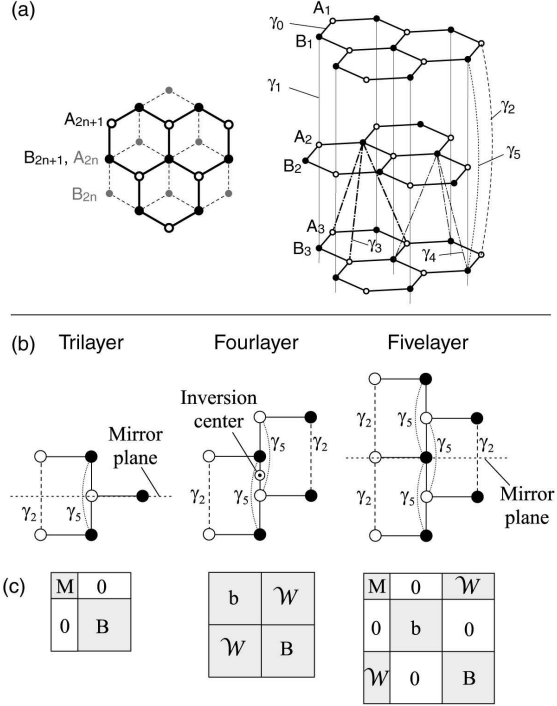}}
\caption{
(a) Atomic structure of ABA-stacked multilayer graphene.
(b) Side view of trilayer (left), fourlayer (center)
and fivelayer (right) lattices,
showing the mirror planes for odd-$N$ layers
and the inversion center for even-$N$ layers.
Horizontal solid lines indicate $\gamma_0$ intralayer coupling between
$A$ (white) and $B$ (black) atoms, vertical solid lines represent
$\gamma_1$. (c)
Schematic form of the partially diagonalized Hamiltonians with
$M$, $b$, $B$ labeling monolayerlike, light bilayerlike, and heavy bilayerlike
blocks, ${\cal W}$ represents coupling between them arising from
next-nearest layer couplings, $\gamma_2$ and $\gamma_5$.
}
\label{fig:1}
\end{figure}
In graphene multilayers with three or more layers, $N \geq 3$ where $N$ is the
number of layers, the effective mass model is much more complicated than
in monolayer or bilayer graphene \cite{mcc06a,guinea06},
or, even, bulk graphite \cite{dressel02}. On the one hand, there are
more relevant parameters than in monolayer or bilayer because of the
presence of {\it next-nearest} layer couplings
while, on the other hand, a lack of translational invariance in
the direction of layer stacking means that the number of basis states in
the model is $2N$, not four as in the
Slonczewski-Weiss-McClure model of bulk graphite \cite{dressel02}.
Although multilayers with a moderate number of layers,
$N \agt 10$, are thought to be similar to bulk graphite \cite{part06},
the properties of few-layer graphene (typically $N \agt 3$) are distinct
\cite{lu06,latil06,part06,guinea06,kosh_mlg,aoki07,min08,kosh09a,kosh09_ssc,avet,avet2,kosh10a,ohta07,guett08,crac08,zhu09,heinz,naka08a,naka08b}.

Here, we consider $ABA$-stacked (Bernal) multilayer graphene,
and analyze its Landau level spectrum and quantum Hall conductivity
in magnetic fields.
In the lattice structure in Fig.~\ref{fig:1}(a) and (b),
the layers, each having two inequivalent atomic sites, $A$ and $B$,
on a honeycomb lattice, are stacked so that half of the atomic sites
have a counterpart directly above or below in the adjacent layers, referred
to as dimer sites, whereas half the sites do not have such a partner (non-dimer
sites).
The effective mass model is characterized by
intralayer coupling $\gamma_0$,
nearest interlayer couplings $\gamma_1$, $\gamma_3$ and $\gamma_4$,
and next-nearest layer couplings $\gamma_2$ and $\gamma_5$
\cite{wallace,slonweiss,dressel02,mcc06a,guinea06}.
By comparison with experiments, it has been possible to determine
the values of relevant parameters including
intralayer coupling \cite{novo05,zhang05,jiang07,martin07} and, in bilayers,
interlayer couplings \cite{ohta06,yan08,zhangli08,li09,mak09,kuz09,hen10}.

While it is possible to numerically diagonalize an effective Hamiltonian
with $2N$ components \cite{lu06,latil06,part06}, this doesn't always
shed light on the roles of the different parameters. Nevertheless,
it was noticed that the multilayer bands form groups of
bilayer-like parabolic bands
near the corners of the Brillouin zone called valleys \cite{kpoints} with,
in odd-$N$ layers, an additional pair of monolayerlike bands
with linear dispersion \cite{guinea06,part06}.
Subsequently, it was realized that a partial diagonalization of the
effective Hamiltonian could be performed in order to write it in block diagonal
form \cite{kosh_mlg,min08,kosh09_ssc}, as illustrated in
Fig.~\ref{fig:1} (c). Each block on the diagonal describes four bilayerlike bands with a given effective
mass (labeled `b' or `B') or, in odd-$N$ layers, two monolayerlike bands
(labeled `M').

The decomposition into block diagonal form is
based upon the eigenstates of a one-dimensional tight-binding chain
in the stacking direction (perpendicular to the layers)
with nearest-neighbor hopping \cite{guinea06}.
If the next-nearest layer couplings, $\gamma_2$ and $\gamma_5$,
are neglected, the decomposition is exact: the diagonal blocks (and their
corresponding bands) are completely separate from each other.
The matrix elements associated with the next-nearest layer couplings
appear within each block
of the decomposed Hamiltonian, but they also couple
separate blocks [as shown by the ${\cal W}$ blocks in Fig.~\ref{fig:1}(c)]
and tend to hybridize their bands \cite{kosh09_ssc}.
In this paper we calculate the Landau levels
with the band parameters fully included,
and show with the aid of decomposition that the next-nearest neighbor couplings
generally account for the energy shifts, the level anti-crossings,
and valley splitting of the low-lying Landau levels,
which significantly influence the phase diagram of the
quantum Hall conductivity.
We focus on trilayer, fourlayer and fivelayer, Fig.~\ref{fig:1},
as representative examples.

\section{The effective-mass model of multilayer graphene}

\subsection{The effective-mass Hamiltonian}

To describe the electronic properties of Bernal-stacked multilayer
graphene, we use an effective-mass model with the
Slonczewski-Weiss-McClure parameterization of graphite \cite{dressel02}.
The low energy spectrum is given by states
in the vicinity of the $K_\xi$ point in the Brillouin zone,
where $\xi = \pm 1$ is the valley index.
If $|A_j\rangle$ and $|B_j\rangle$ are Bloch functions at the $K_{\xi}$
point, corresponding to the $A$ and $B$ sublattices of layer $j$,
respectively, then a suitable basis is $|A_1\rangle,|B_1\rangle$;
$|A_2\rangle,|B_2\rangle$; $\cdots$; $|A_N\rangle,|B_N\rangle$.
In this basis, the Hamiltonian of multilayer graphene with $N$ layers
\cite{guinea06,part06,lu06,kosh_mlg,kosh09_ssc}
in the vicinity of the $K_{\xi}$ valley is
\begin{eqnarray}
H_{N} =
\begin{pmatrix}
 H_0 & V & W & & \\
 V^{\dagger} & H_0' & V^{\dagger}& W'& \\
 W & V & H_0 & V & W & \\
& W' & V^{\dagger} & H_0' & V^{\dagger} & W' & \\
  & &  \ddots & \ddots & \ddots & \ddots & \ddots
\end{pmatrix},
\label{eq_H}
\end{eqnarray}
with
\begin{eqnarray}
&& H_0 =
\begin{pmatrix}
 0 & v \pi^\dagger \\ v \pi & \Delta'
\end{pmatrix},
\quad
H_0' =
\begin{pmatrix}
 \Delta' & v \pi^\dagger \\ v \pi & 0
\end{pmatrix},
\label{Hdef}
\\
&& V =
\begin{pmatrix}
 -v_4\pi^\dagger & v_3 \pi \\ \gamma_1 & -v_4\pi^\dagger
\end{pmatrix} , \label{Vdef}
\\
&& W =
\begin{pmatrix}
 \gamma_2/2 & 0 \\ 0 & \gamma_5/2
\end{pmatrix} , \quad
 W' =
\begin{pmatrix}
 \gamma_5/2 & 0 \\ 0 & \gamma_2/2
\end{pmatrix} .
\label{Wdef}
\end{eqnarray}
Here, the in-plane momentum operator is $\pi = \xi p_x + i p_y$,
and $\GVec{p} = \left( p_x , p_y \right) = -i\hbar \nabla + e \Vec{A}$
with vector potential $\Vec{A}$.
The diagonal blocks, Eq.~(\ref{Hdef}), describe nearest-neighbor intralayer
coupling, and $V$, Eq.~(\ref{Vdef}), describes nearest-neighbor layer
coupling, where $\gamma_1$ is the interlayer coupling between dimer sites.
Parameter $\Delta'$ represents the energy difference
between dimer sites and non-dimer sites, and thus it only exists for $N \geq 2$.
It is related to the band parameters as
$\Delta' = \Delta -\gamma_2 + \gamma_5$.
The Fermi velocity of monolayer graphene is $v = \sqrt{3} a \gamma_0/2\hbar$,
and other velocities are defined as $v_3 = \sqrt{3} a \gamma_3/2\hbar$
and $v_4 = \sqrt{3} a \gamma_4/2\hbar$.
Matrix $W$, Eq.~(\ref{Wdef}), describes coupling between next-nearest
neighboring layers, and it only exists for $N \geq 3$. Parameters
$\gamma_2$ and $\gamma_5$ couple a pair of non-dimer sites and
a pair of dimer sites, respectively.

\subsection{Decomposition to an approximate block diagonal form}

The decomposition of the Hamiltonian, Eq. (\ref{eq_H}), into
an approximate block diagonal form uses a unitary transformation based on the
eigenstates of a linear chain of atoms in the $z$-direction
\cite{kosh_mlg,min08,kosh09_ssc}:
\begin{eqnarray}
 f_m(j) =  c_m \sqrt{\frac{2}{N+1}}[1 - (-1)^j] \sin \kappa_m j \\
 g_m(j) =  c_m \sqrt{\frac{2}{N+1}}[1 + (-1)^j] \sin \kappa_m j,
\end{eqnarray}
where
\begin{eqnarray}
&&\kappa_m = \frac{\pi}{2}-\frac{m\pi}{2(N+1)},
 \label{eq_lambda}\\
&&
c_m = \left\{
\begin{array}{ll}
1/2 & (m=0) \\
1/\sqrt{2} & (m\neq 0).
\end{array}
\right.
\end{eqnarray}
Here $j=1,2,\cdots,N$ is the layer index,
and $m$ is the block index which ranges as
\begin{eqnarray}
 m =
\left\{
\begin{array}{l}
1,3,5, \cdots , N-1, \quad N = {\rm even} \, , \\
0,2,4, \cdots , N-1, \quad N = {\rm odd} \, .
\end{array}
\right.
\label{eq_m}
\end{eqnarray}
Obviously $f_m(j)$ is zero on even $j$ layers, while $g_m(j)$ is zero on odd $j$ layers.
The basis is constructed \cite{kosh_mlg,kosh09_ssc} by assigning $f_m(j)$, $g_m(j)$
to each site as
\begin{eqnarray}
 |\phi_{m}^{\rm (X, odd)}\rangle \!\! &=& \!\!
\sum_{j=1}^N f_m(j) |X_j\rangle \, , \nonumber\\
 |\phi_{m}^{\rm (X, even)}\rangle \!\! &=& \!\!
\sum_{j=1}^N g_m(j) |X_j\rangle \, ,
\label{eq_basis}
\end{eqnarray}
where $X=A$ or $B$. A superscript such as (A, odd) indicates that
the wave function has a non-zero amplitude only on $|A_j\rangle$ sites with odd $j$'s.

In order to write the Hamiltonian Eq.~(\ref{eq_H})
in terms of the basis states Eq.~(\ref{eq_basis}),
we group the basis of block $m$ as
$\Vec{u}_m
= \{|\phi_m^{\rm (A, odd)}\rangle, \, |\phi_{m}^{\rm (B, odd)}\rangle, \,
|\phi_m^{\rm (A, even)}\rangle, \, |\phi_{m}^{\rm (B, even)}\rangle\}$.
Then, the block matrix between different $m$'s may be written as
\begin{eqnarray}
{\cal H}_{m'm} \!\equiv\! \Vec{u}^\dagger_{m'} H \Vec{u}_m
\!=\! {\cal H}(\lambda_m)\delta_{m'm} + {\cal W}(\alpha_{m'm},\beta_{m'm}),
\label{genblocks}
\end{eqnarray}
with
\begin{eqnarray}
&&\!\!\!\!\!\! \!\!\!\!\!\! {\cal H}(\lambda) =
\begin{pmatrix}
0 & v\pi^{\dagger} &  -\lambda v_4\pi^{\dagger} & \lambda v_3\pi \\
v\pi & \Delta'
& \lambda \gamma_1 &  -\lambda v_4\pi^{\dagger} \\
 -\lambda v_4\pi &  \lambda \gamma_1 & \Delta'
& v\pi^{\dagger} \\
 \lambda v_3\pi^{\dagger} &  -\lambda v_4\pi & v\pi & 0
\end{pmatrix} ,
\label{eq_Hm}
\\
&&{\cal W}(\alpha,\beta) =
\begin{pmatrix}
\alpha \gamma_2 & 0 & 0 & 0 \\
0 & \alpha \gamma_5  & 0 & 0 \\
0 & 0 & \beta \gamma_5 & 0 \\
0 & 0 & 0 &  \beta \gamma_2
\end{pmatrix},
\label{eq_Wm}
\end{eqnarray}
where
\begin{eqnarray}
\lambda_m &=& 2 \cos \kappa_m, \\
\alpha_{m'm} &=& 2c_mc_{m'}\Biggl\{
\delta_{mm'}(1+\delta_{m0}) \cos 2\kappa_m +
\nonumber\\
&& \frac{\sin\kappa_m \sin\kappa_{m'}}{N+1}
\left[
2+(-1)^{\frac{m-m'}{2}}(1-(-1)^N)
\right]
\Biggr\}, \nonumber\\
\\
\beta_{m'm} &=&  2c_mc_{m'}\Biggl\{
\delta_{mm'}(1-\delta_{m0}) \cos 2\kappa_m +
\nonumber\\
&&
\frac{\sin\kappa_m \sin\kappa_{m'}}{N+1}
(-1)^{\frac{m-m'}{2}}(1+(-1)^N)
\Biggr\}.
\end{eqnarray}
The diagonal matrix ${\cal H}_{mm}$
is equivalent to the Hamiltonian of bilayer graphene
\cite{mcc06a} with nearest-layer coupling parameters
multiplied by $\lambda$
\cite{guinea06,part06,kosh_mlg,min08,kosh09_ssc},
and on-site asymmetric potential described by ${\cal W}_{mm}$.
The off-diagonal block, ${\cal W}$ for $m \neq m'$, has not been
explicitly obtained before: it appears only when
coupling between the next-nearest neighboring layers,
$\gamma_2$ and $\gamma_5$, is non-zero.
The block ${\cal W}$ is diagonal,
where $\gamma_2$ only connects pairs of non-dimer sites,
and $\gamma_5$ only connects pairs of dimer sites,
of these effective bilayer-like blocks.

The case of $m=0$ is special in that
$g_m(j)$ is identically zero, so that
only two basis states
$\{|\phi_0^{\rm (A, odd)}\rangle, \, |\phi_{0}^{\rm (B, odd)}\rangle\}$
survive in Eq.~(\ref{eq_basis}).
The matrix elements associated with the two missing basis states should be neglected
in Eqs.~(\ref{eq_Hm},\ref{eq_Wm}).
Specifically, the matrix for the $m=0$ block written in the two component basis
is
\cite{kosh09_ssc}
\begin{eqnarray}
{\cal H}_{0} =
\begin{pmatrix}
0 & v\pi^\dagger \\
v\pi &  \Delta'
\end{pmatrix}
- \frac{N-1}{N+1}
\begin{pmatrix}
 \gamma_2 & 0 \\
0 &   \gamma_5
\end{pmatrix} ,
\label{eq_H0}
\end{eqnarray}
which, barring the diagonal terms,
is equivalent to the Hamiltonian of monolayer graphene.


We stress the role of the symmetry of the lattice
and note that the even-odd effect, with respect to the number of layers,
goes further than the absence or presence of monolayerlike bands in
the band structure. The lattice of odd-$N$ multilayers obeys
mirror reflection symmetry
$(x,y,z) \rightarrow ( x, y , -z)$
\cite{latil06,manes07,kosh_mlg,kosh09a,kosh10a}
(mirror planes for trilayer and fivelayer graphene
are shown in Fig.~\ref{fig:1}(b)),
and thus the eigenstates can be classified by
parity with respect to the reflection:
the parity of the wavefunction of the subband $m$
is given by $(-1)^{\frac{N-m-1}{2}}$,
so the group of $m=2,6,10,\ldots$
and that of $m=0,4,8,\ldots$ have opposite parities.
Since eigenstates with different parity cannot be mixed
by terms in the Hamiltonian that preserve lattice symmetry,
off-diagonal blocks connecting diagonal blocks with different parity
are identically zero
even in the presence of next-nearest layer couplings \cite{kosh09_ssc,kosh09a}.
We actually see that the coupling matrix ${\cal W}_{mm'}$
between blocks having different parities indeed vanishes,
as illustrated in Fig.~\ref{fig:1}(c) for trilayer and fivelayer.

Even-$N$ multilayers lack mirror reflection symmetry, however, so that
$\gamma_2$ and $\gamma_5$ mix every diagonal block, as shown
in Fig.~\ref{fig:1}(c)
for fourlayer graphene. Instead, the lattice of even-$N$ multilayers
obeys spatial inversion symmetry
$(x,y,z) \rightarrow (- x, -y , -z)$ \cite{latil06,manes07,kosh_mlg,kosh09a,kosh10a}
(an inversion center for fourlayer graphene is shown in Fig.~\ref{fig:1}, center).
Unlike mirror reflection, inversion symmetry transforms electronic states between valleys,
and, even in the presence of a magnetic field, this ensures degeneracy
of the electronic spectra at different valleys \cite{kosh10a}.

\subsection{Reduced low-energy Hamiltonian}

As we show below, mixing between blocks is particularly important
in the vicinity of level crossings. Level crossings apart, a good approximation
to the spectra over a broad range of energy may be obtained by
neglecting the off-diagonal blocks \cite{kosh09_ssc}.
Taken alone, the bilayerlike block ${\cal H}_{mm}$ for $m\neq 0$
describes four bands \cite{mcc06a},
two split off by energy $\pm \lambda \gamma_1$
at the $K_{\xi}$ point and two near zero energy.
The split bands can be viewed as a bonding and anti-bonding pair
created by the relatively strong interlayer coupling $\gamma_1$ between
dimer sites (A, even) and (B, odd).
For low energy, $\epsilon \ll \lambda \gamma_1$, it is possible to
derive a reduced Hamiltonian for the bilayerlike block describing
an effective hopping between non-dimer sites (A, odd) and (B, even)
by using a Schrieffer-Wolff transformation \cite{swtrans,mcc06a}
to eliminate components $|\phi_m^{\rm (A, even)}\rangle$
and $|\phi_m^{\rm (B, odd)}\rangle$.
Then, the basis of block $m \neq 0$ is reduced to
$\widetilde{\Vec{u}}_m
= \{|\phi_m^{\rm (A, odd)}\rangle, |\phi_{m}^{\rm (B, even)}\rangle \}$
and the Hamiltonian matrix for the block is
modified as
\begin{eqnarray}
\widetilde{\cal H}_{mm}
&=& \widetilde{\cal H}(\lambda_m)
+ \widetilde{\cal W}(\alpha_{mm},\beta_{mm}),
\\
\widetilde{{\cal H}} (\lambda) &=&
- \frac{v^2}{\lambda \gamma_1}\left(
  \begin{array}{cc}
    0 & \left( \pi^{\dagger} \right)^2 \\
    \pi^2 & 0 \\
  \end{array}
\right) +
\lambda v_3 \left(
  \begin{array}{cc}
    0 & \pi \\
    \pi^{\dagger} & 0 \\
  \end{array}
\right)
\nonumber\\
&& + \frac{2vv_4}{\gamma_1}
\left(
  \begin{array}{cc}
    \pi^\dagger\pi & 0 \\
    0 & \pi\pi^{\dagger}  \\
  \end{array}
\right)
, \label{eq_Hmr} \\
\!\!\!\!\!\!\!\!\!\!\!\!\widetilde{{\cal W}} (\alpha,\beta) &=&
\left(
  \begin{array}{cc}
    \alpha \gamma_2 & 0 \\
    0 & \beta \gamma_2 \\
  \end{array}
\right) . \label{eq_Wmr}
\end{eqnarray}
This reduced Hamiltonian is approximately valid
at low energy $\{  \epsilon, vp \} \ll \lambda \gamma_1$.

When $v_3$ and $v_4$ are neglected, the eigenvalues of
$\widetilde{\cal H}_{mm}$ at zero magnetic field are
\begin{eqnarray}
\epsilon^{(m)}_{\pm}(p) \!=\! \frac{\alpha+\beta}{2}\gamma_2
\pm \sqrt{
\left(\frac{\alpha-\beta}{2}\gamma_2\right)^2 +
\frac{v^4p^4}{(\lambda\gamma_1)^2}},
\label{eq_disp_m}
\end{eqnarray}
with $\lambda = \lambda_m$, $\alpha = \alpha_{mm}$, $\beta = \beta_{mm}$.
This gives nearly-parabolic conduction and valence bands
centered at energy $(\alpha+\beta)\gamma_2/2$
with an energy gap $|(\alpha-\beta)\gamma_2|$ between them.
The gap always vanishes for even-layered graphene
since $\alpha_{mm} = \beta_{mm}$ holds for all $m$
when $N$ is even.
The extra parameter $v_3$ introduces trigonal warping
in a similar manner as in bilayer graphene \cite{mcc06a},
and the $v_4$ parameter produces a weak electron-hole asymmetry
by adding the band energy $2vv_4p^2/\gamma_1$ in both the
conduction and valence bands \cite{kosh09b}.

The Landau level spectrum
in a uniform and perpendicular magnetic field may be found using the
Landau gauge $\mathbf{A} = (0, Bx , 0)$. Then, at valley $K_{+}$,
the operators $\pi$ and $\pi^{\dag }$ coincide with raising
and lowering operators \cite{p+r87} in the basis of Landau
functions $\psi_n (x,y) = e^{ip_yy/\hbar}\phi_{n}(x - p_y \lambda_B^2)$,
such that
$\pi\psi_{n}=i(\hbar /\lambda_{B})\sqrt{2(n+1)}\psi_{n+1}$,
$\pi^{\dag } \,\psi_{n}=-i(\hbar /\lambda _{B})\sqrt{2n}\psi_{n-1}$,
and $\pi^{\dag } \,\psi_{0}=0$.
Here $\lambda_B = \sqrt{\hbar/(eB)}$ is the magnetic length.
At valley $K_{-}$, the effect of the operators
becomes
$\pi^{\dag } \,\psi_{n}=-i(\hbar /\lambda_{B})\sqrt{2(n+1)}\psi_{n+1}$,
$\pi\psi_{n}=i(\hbar /\lambda _{B})\sqrt{2n}\psi_{n-1}$,
and $\pi\psi_{0}=0$.
In the absence of $v_3$ and $v_4$,
the Landau level spectrum for the Hamiltonian $\widetilde{\cal H}_{mm}$
at the valley $\xi$ is given by
\begin{eqnarray}
\epsilon_{n\geq 1,\pm}^{(m)} &=&
 \frac{\alpha+\beta}{2}\gamma_2
\pm \sqrt{
\left(\frac{\alpha-\beta}{2}\gamma_2\right)^2 +
\frac{n \left(n + 1\right) \Gamma_B^4}{(\lambda\gamma_1)^2}},
\nonumber\\
\epsilon_{n=-1}^{(m)} &=& \epsilon_{n=0}^{(m)} =
\left(
\frac{1 + \xi}{2} \alpha
+ \frac{1 - \xi}{2} \beta \right) \gamma_2,
\label{eq_landau_m}
\end{eqnarray}
where $\Gamma_B = \sqrt{2\hbar v^2 eB} = \sqrt{2} \hbar v / \lambda_B$
and we consider $\{ |\epsilon|, \sqrt{n} \Gamma_B \} \ll |\gamma_1|$.
When $\alpha \neq \beta$, each of two lowest levels at $n=-1,0$
split in valleys due to next-layer coupling $\gamma_2$,
moving to energy corresponding to either
the bottom of the zero-field conduction band or the top of the valence band.
Other levels are valley degenerate in this approximation.

Even the degeneracy of the $n=0$ and $n=-1$ levels is lifted 
in the higher order of $\Gamma_B/(\lambda \gamma_1)$.
By applying a perturbation
to the original $4\times 4$ Hamiltonian, we find the correction
to be
\begin{eqnarray}
&&\delta \epsilon_{n=-1}^{(m)} = 0,
\nonumber\\
&&\delta \epsilon_{n=0}^{(m)} =
\left[
\left(
\frac{1 + \xi}{2} \alpha
+ \frac{1 - \xi}{2} \beta \right)
\gamma_5
+\Delta'
+ \frac{2\lambda v_4}{v}\lambda\gamma_1
\right]
\nonumber\\
&& \hspace{50mm} \times\frac{\Gamma_B^2}{(\lambda\gamma_1)^2},
\label{eq_split}
\end{eqnarray}
so that the splitting is proportional to $B$.


The monolayerlike block for $m=0$, Eq.\ (\ref{eq_H0}),
is characterized by only one parameter with $\alpha=-(N-1)/(N+1)$.
The energy dispersion is given by
\begin{eqnarray}
\epsilon_{\pm}^{(0)}(p) &=& \frac{\Delta'+\alpha(\gamma_2 + \gamma_5)}{2}
\nonumber\\
&&
\pm \sqrt{\left(
\frac{\Delta'-\alpha(\gamma_2 - \gamma_5)}{2} \right)^2 + v^2p^2 },
\label{eq_disp_0}
\end{eqnarray}
which generally has an energy gap
of the width $|\Delta'-\alpha(\gamma_2 - \gamma_5)|$
at Dirac point, and an overall energy shift of
$[\Delta'+\alpha(\gamma_2 + \gamma_5)]/2$.
In a magnetic field, Landau levels become \cite{kosh10}
\begin{eqnarray}
\epsilon_{n\geq 1,\pm}^{(0)} &=&
\frac{\Delta'+\alpha(\gamma_2 + \gamma_5)}{2}
\nonumber\\
&&\pm \sqrt{
\left(
\frac{\Delta'-\alpha(\gamma_2 - \gamma_5)}{2} \right)^2
+ n \Gamma_B^2 } , \nonumber\\
\epsilon_{n=0}^{(0)} &=&
\frac{1 + \xi}{2}\alpha\gamma_2
+
\frac{1 - \xi}{2}(\Delta'+\alpha\gamma_5).
\label{eq_landau_0}
\end{eqnarray}
Similarly to the bilayer-like block, the lowest levels at $n=0$
of $K_+$ and $K_-$ split
to the energy of the bottom of the zero-field conduction band
or the top of the valence band, while
the other levels are valley degenerate.

The Landau levels of the different blocks
are hybridized by the off-diagonal matrix ${\cal W}_{mm'}$.
At the valley $K_+$
the wavefunction of the Landau level with index $n$
can be written in form
$(c_1 \psi_{n-1}, c_2 \psi_n, c_3\psi_n, c_4\psi_{n+1})$
for the bilayer-like band, and
$(c_1 \psi_{n-1}, c_2 \psi_{n})$ for the monolayer-like band.
Since ${\cal W}_{mm'}$ is diagonal and does not
include $\pi$ or $\pi^\dagger$, it only couples Landau levels
of different blocks if they have the same index $n$.
Furthermore, parameter $\gamma_3$  mixes
levels $n$ and $n+3$ within the same block,
and thus it, together with ${\cal W}_{mm'}$, leads to hybridization
among the levels of different blocks
whose indices are equal in modulo 3.
Such coupling leads to anticrossing at
the intersecting point of the corresponding Landau levels.
At the valley $K_-$,
the wavefunction of the Landau level with index $n$ becomes
$(c_1 \psi_{n+1}, c_2 \psi_n, c_3\psi_n, c_4\psi_{n-1})$
for the bilayer-like band, and
$(c_1 \psi_{n}, c_2 \psi_{n-1})$ for the monolayer-like band,
where the index at the same position differs between monolayer
and bilayer.
As a result, the above rule for $K_+$ changes
only for the coupling between monolayer and bilayer levels,
where the $n$-th monolayer level couples with the $n'$-th bilayer
level when $n-1$ and $n'$ are equal in modulo 3.

\section{Trilayer graphene}

According to the decomposition described previously,
the Hamiltonian in basis
$|\phi_0^{\rm (A, odd)}\rangle, \, |\phi_{0}^{\rm (B, odd)}\rangle, $
$|\phi_2^{\rm (A, odd)}\rangle, \, |\phi_{2}^{\rm (B, odd)}\rangle, \,
|\phi_2^{\rm (A, even)}\rangle, \, |\phi_{2}^{\rm (B, even)}\rangle$
may be written in block diagonal form \cite{kosh09a} as
\begin{eqnarray}
H_{N=3} &=& \left(
\begin{array}{cc}
  {\cal H}_{0} & 0 \\
  0 & {\cal H}_{2} \\
\end{array}
\right) , \label{h3}
\end{eqnarray}
where
\begin{eqnarray}
{\cal H}_{0} &=&
\begin{pmatrix}
0 & v\pi^\dagger \\
v\pi &  \Delta'
\end{pmatrix}
- \frac{1}{2}
\begin{pmatrix}
 \gamma_2 & 0 \\
0 &   \gamma_5
\end{pmatrix} ,
\label{hm} \\
{\cal H}_{2} &=&  {\cal H}\left( \lambda_2 \right) + {\cal W}(1/2, 0),
\label{hb}
\end{eqnarray}
where $\lambda_2 = \sqrt{2}$.
The off-diagonal blocks in Eq.~(\ref{h3}) connecting
the monolayerlike block ${\cal H}_{0}$ and the
bilayerlike block ${\cal H}_{2}$ are identically zero,
because the basis states for
the monolayerlike and bilayerlike blocks have different
parity with respect to mirror reflection as argued above.

In absence of $v_3$ and $v_4$,
the low-energy Landau level spectrum
for the bilayer-like band $(m=2)$ is given by Eq.\ (\ref{eq_landau_m})
with $(\lambda, \alpha,\beta) =(\sqrt{2},1/2,0)$,
and that for the monolayer-like band $(m=0)$ by
Eq.\ (\ref{eq_landau_0}) with $\alpha=-1/2$.
The lowest Landau levels of each block
$\epsilon_{n=0}^{(0)}$, $\epsilon_{n=-1}^{(2)}$ and
$\epsilon_{n=0}^{(2)}$,
which are degenerate in the absence of next-layer coupling \cite{guinea06},
split according to
\begin{eqnarray}
\epsilon_{n=0}^{(0)} &=&
\frac{1+\xi}{2}\left(-\frac{\gamma_2}{2}\right)
+ \frac{1-\xi}{2} \left( \Delta' - \frac{\gamma_5}{2} \right), \label{eq_zero_3a} \\
\epsilon_{n=-1}^{(2)} &=& \epsilon_{n=0}^{(2)} = \left( 1 + \xi \right)
 \frac{\gamma_2}{4}.
\label{eq_zero_3}
\end{eqnarray}
The two lowest levels of the bilayerlike block
$\epsilon_{n=-1}^{(2)}, \epsilon_{n=0}^{(2)}$,
are weakly split by extra band parameters
in accordance with Eq.\ (\ref{eq_split}).
As a result, the twelve zero-energy levels
split into six different energies,
each of them having twofold spin degeneracy.

We numerically calculate the Landau level spectrum
by diagonalizing the original Hamiltonian Eq.\ (\ref{eq_H})
including all parameters.
We adopt the parameter values \cite{dressel02}
$\gamma_2 = -0.02$ eV, $\gamma_5 = 0.04$ eV,
$\Delta' = 0.05$ eV, $\gamma_0 = 3$ eV,
$\gamma_1 = 0.4$ eV, $\gamma_3 = 0.3$ eV,
$\gamma_4 = 0.04$ eV.
Since the dimension of the Hamiltonian matrix becomes
infinite in the presence of trigonal warping,
we introduce a cut-off in the Landau level index, $n=100$,
which is enough high to obtain the proper low-energy spectrum \cite{kosh09b}.

\begin{figure}[t]
\centerline{\epsfysize=1.6\hsize \epsffile{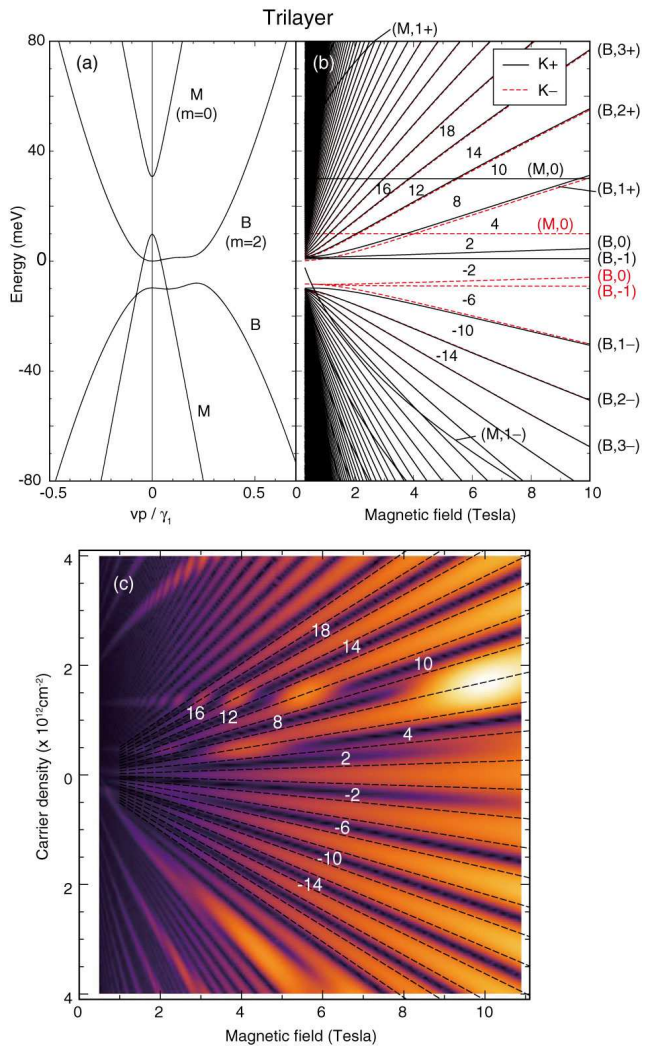}}
\caption{(a) Low-energy band structure and (b) Landau levels
as a function of magnetic field of ABA-trilayer graphene.
(c) Two-dimensional density plot of the density of states
in the space of magnetic field and carrier density.
In (b) and (c), numbers represent the quantized Hall
conductivity in units of $e^2/h$.
}
\label{fig_3lyr}
\end{figure}

Fig.\ \ref{fig_3lyr}(a) and (b) show
the zero-field dispersion and the Landau levels plotted against the
magnetic field, respectively, of ABA trilayer graphene
with all the band parameters included.
The symbols `M' and `B' represent the monolayer
block $(m=0)$ and the bilayer block $(m=2)$, respectively.
The spectrum is composed of monolayerlike and bilayerlike Landau levels
that are shifted relatively to each other in energy,
as qualitatively described by the Hamiltonian decomposition above,
Eqs.~(\ref{h3}-\ref{hb}).
Zero energy levels are close to those obtained analytically, Eqs.\ (\ref{eq_zero_3a},\ref{eq_zero_3}),
corresponding to the bottom of the zero-field conduction band or the top
of the valence band. We also observe the
weak splitting of $\epsilon_{n=-1}^{(2)}, \epsilon_{n=0}^{(2)}$
[indicated as (B,$-1$), (B,0), respectively] argued above.

Fig.\ \ref{fig_3lyr}(c) shows a two-dimensional plot of the
density of states in the space of magnetic field
and the carrier density.
For simplicity, we assume that each Landau level is broadened
into a Gaussian shape with width
$\Delta E = C \Gamma_B$, \cite{ando74,Shon_and_Ando_1998a}
where the constant $C$ is taken to be 0.03.
The dark and bright colors represent low and high density of states.
Apart from the usual Landau fan,
we observe a series of bright lines corresponding to the
crossings of monolayerlike and bilayerlike Landau levels \cite{pablo}.
The numbers assigned to the dark regions 
indicate the quantized Hall conductivity in units of $e^2/h$,
and correspond to those assigned
to spaces between Landau levels in Fig.\ \ref{fig_3lyr}(b).
The quantized Hall conductivity jumps by $2$
only at the zero levels $\epsilon_{n=0}^{(0)}$,
$\epsilon_{n=-1}^{(2)}$, $\epsilon_{n=0}^{(2)}$
which split in valleys,
while it changes in units of 4 otherwise since other levels
are almost valley degenerate.
As a result, the quantized Hall conductivity
becomes $4M$ ($M$: integer) only in the region between
$\epsilon_{n=0}^{(0)}$ levels of two valleys,
corresponding to the energy gap of the monolayer band,
and also in the narrow regions between
$\epsilon_{n=-1}^{(2)}$ and $\epsilon_{n=0}^{(2)}$
for each of $K_+$ and $K_-$.
Otherwise it takes a series of $4M+2$.

\section{Fourlayer graphene}

The Hamiltonian of ABA-stacked fourlayer Hamiltonian,
Eq.~(\ref{eq_H}) for $N=4$, may be partially decomposed into
two bilayerlike blocks with subsystem indices $m=1$ and $m=3$:
\begin{eqnarray}
H_{N=4}
=
      \begin{pmatrix}
      {\cal H}_{1} & {\cal{H}}_{13} \\
      {\cal{H}}_{31} & {\cal H}_{3} \\
     \end{pmatrix}
\end{eqnarray}
where
\begin{eqnarray}
{\cal H}_{1} \!\!&=&\!\!
{\cal H} \left( \lambda_1 \right)
+ {\cal W}(-p, -p), \label{eq_H1} \\
{\cal H}_{3} \!\!&=&\!\!
{\cal H} \left( \lambda_3 \right)
+ {\cal W}(p, p), \label{eq_H3} \\
{\cal{H}}_{13} \!\!&=&\!\! {\cal{H}}_{31} =
{\cal W}(p/2, -p/2).
\end{eqnarray}
and
$\lambda_1 = (-1+\sqrt{5})/2$, 
$\lambda_3 =  (1+\sqrt{5})/2$, 
and $p = 1/\sqrt{5}$.

By neglecting the inter-block mixing
as well as $v_3$ and $v_4$,
the low-energy Landau level spectrum is given by Eq.(\ref{eq_landau_m}),
with $(\lambda,\alpha,\beta)=(\lambda_1,-p,-p)$ for $m=1$,
and $(\lambda_3,p,p)$ for $m=3$.
The zero-energy Landau levels become
\begin{eqnarray}
\epsilon_{n=-1}^{(1)} &=& \epsilon_{n=0}^{(1)} = - \frac{\gamma_2}{\sqrt{5}}
\\
\epsilon_{n=-1}^{(3)} &=& \epsilon_{n=0}^{(3)} = \frac{\gamma_2}{\sqrt{5}},
\end{eqnarray}
which are valley degenerate as they should be, due to
spatial inversion symmetry.
The two lowest levels of each bilayer-like
block, $\epsilon_{n=-1}^{(m)}, \epsilon_{n=0}^{(m)}$,
are split due to extra parameters as described by Eq.\ (\ref{eq_split}),
so that the sixteen zero-energy levels
split into four energies, each of them with fourfold
spin and valley degeneracy retained.

\begin{figure}[t]
\centerline{\epsfysize=1.6\hsize \epsffile{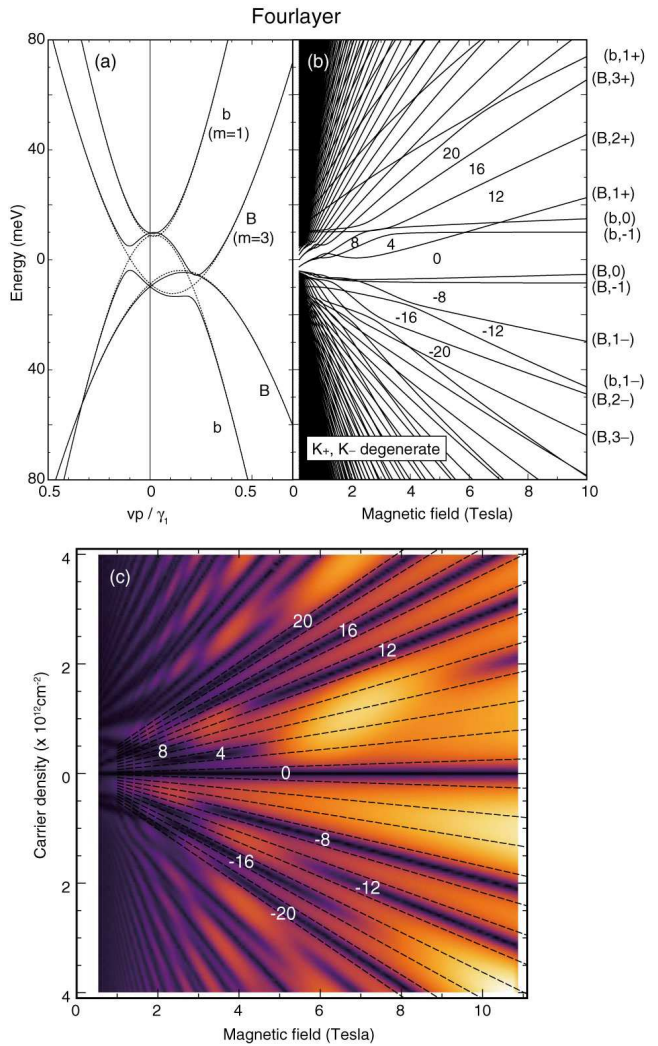}}
\caption{Plots similar to Fig.\ \ref{fig_3lyr}
for ABA fourlayer graphene.
In (a), the dotted curve indicates energy bands calculated
by neglecting the off-diagonal block.
}
\label{fig_4lyr}
\end{figure}

Fig.\ \ref{fig_4lyr} (a) and (b) show
the zero-field band structure, and
the  Landau levels of fourlayer graphene, respectively
computed numerically including all parameters.
In (a), the dotted curve indicates the energy bands with the
off-diagonal block neglected.
The symbols `b' and `B' represent the light-mass bilayer
$(m=1)$ and the heavy-mass bilayer block $(m=3)$, respectively.
The Landau level spectrum is basically regarded as composition
of two series of bilayerlike levels,
while there are anticrossings caused by ${\cal H}_{13}$
between the levels having the same index in modulo 3.
The valley degeneracy is never broken even in the full parameter model
as it is protected by the spatial inversion symmetry of the lattice.

Fig.\ \ref{fig_4lyr}(c) illustrates a
two-dimensional map of the density of states
calculated from the Landau levels in Fig.\ \ref{fig_4lyr}(b).
The quantum Hall integer
is always $4M$ ($M$: integer) because all the levels are
valley and spin degenerate.
Similarly to the trilayer characteristic bright lines
appear at crossing points of Landau levels
belonging to the different bilayer blocks.

\section{Fivelayer graphene}

The Hamiltonian of ABA-stacked fivelayer Hamiltonian,
Eq.~(\ref{eq_H}) for $N=5$, may be partially decomposed into
a monolayerlike block $m=0$ and
two bilayerlike blocks with subsystem indices $m=2$ and $m=4$:
\begin{eqnarray}
H_{N=5}=
      \begin{pmatrix}
      {\cal H}_{0} & 0 & {\cal{H}}_{04} \\
      0 & {\cal{H}}_{2} & 0 \\
      {\cal H}_{40} & 0 & {\cal{H}}_{4} \\
     \end{pmatrix}
\end{eqnarray}
where
\begin{eqnarray}
{\cal H}_{0} &=&
\begin{pmatrix}
0 & v\pi^\dagger \\
v\pi &  \Delta'
\end{pmatrix}
- \frac{2}{3}
\begin{pmatrix}
 \gamma_2 & 0 \\
0 &   \gamma_5
\end{pmatrix} ,
\\
{\cal H}_{2} \!\!&=&\!\!
{\cal H} \left(\lambda_2 \right)
+ {\cal W}(0, -1/2), 
\\
{\cal H}_{4} \!\!&=&\!\!
{\cal H} \left(\lambda_4 \right)
+ {\cal W}(2/3, 1/2), 
\\
{\cal{H}}_{04} \!\!&=&\!\! {\cal{H}}_{40}^\dagger =
\bigl[{\cal W}(\sqrt{2}/6, 0)\bigl]_{2\times 4},
\end{eqnarray}
where $\lambda_2 = 1$, $\lambda_4 =  \sqrt{3}$,
and $[{\cal W}]_{2\times 4}$ represents the upper half
(i.e., the first two rows) of the matrix ${\cal W}$.
The block $m=2$ is never mixed with $m=0$ and 4,
due to the parity difference as argued previously.

By neglecting the inter-block mixing, $v_3$ and $v_4$,
the zero-energy Landau levels are given by
\begin{eqnarray}
\epsilon_{n=0}^{(0)} &=&
\frac{1+\xi}{2}\left(-\frac{\gamma_2}{2}\right)
+ \frac{1-\xi}{2} \left( \Delta' - \frac{\gamma_5}{2} \right), \\
\epsilon_{n=-1}^{(2)} &=& \epsilon_{n=0}^{(2)} =
-(1-\xi)\gamma_2\\
\epsilon_{n=-1}^{(4)} &=& \epsilon_{n=0}^{(4)} =
\left(
\frac{1 + \xi}{3}+ \frac{1 - \xi}{4} \right) \gamma_2,
\label{eq_zero_5}
\end{eqnarray}
Similarly to the trilayer and fourlayer,
the two lowest levels of each bilayer-like
block, $\epsilon_{n=-1}^{(m)}, \epsilon_{n=0}^{(m)}$,
split due to extra parameters in accordance with Eq.\ (\ref{eq_split}),
resulting in ten different zero-energy levels with spin degeneracy.

\begin{figure}[t]
\centerline{\epsfysize=1.6\hsize \epsffile{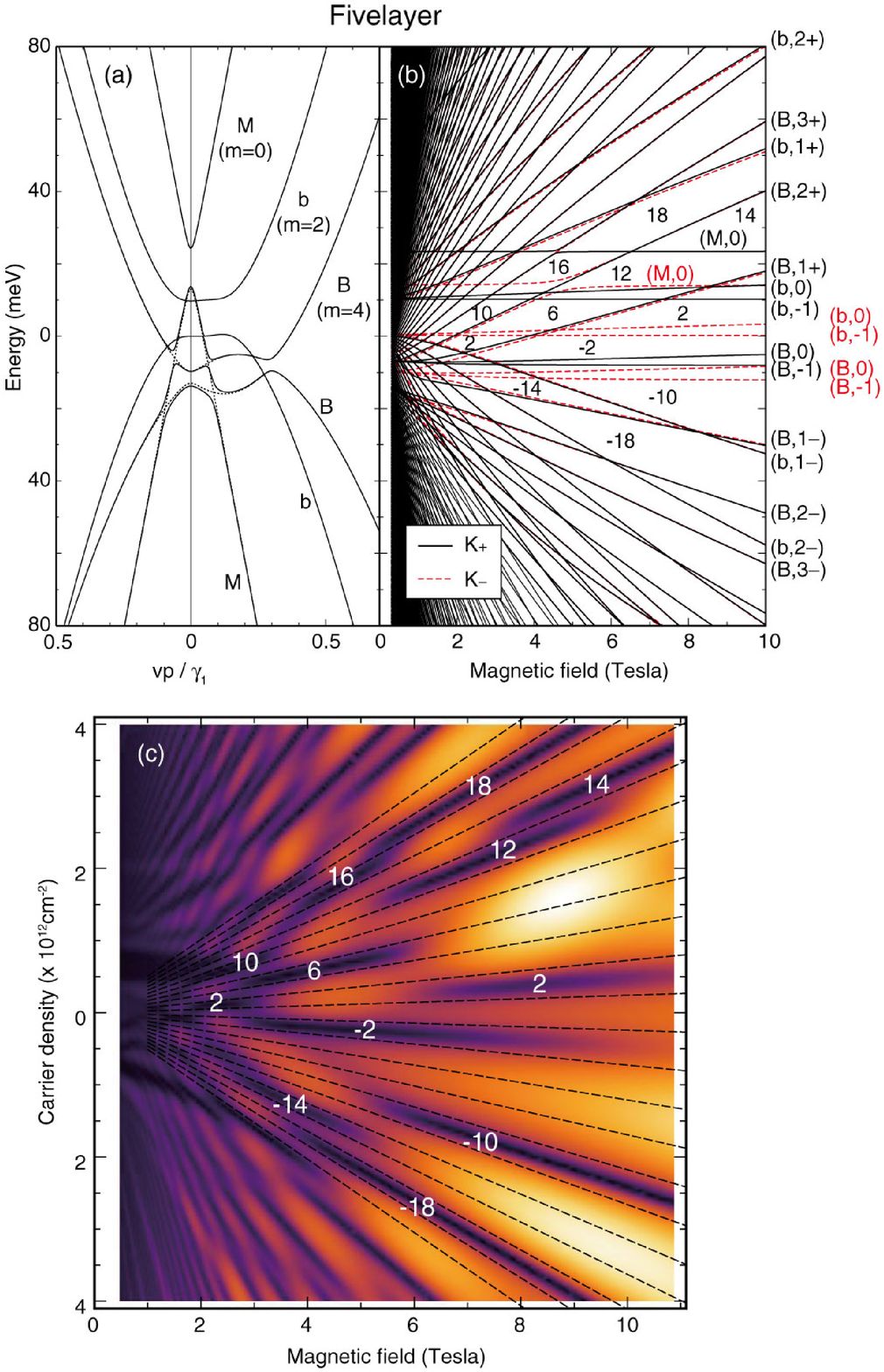}}
\caption{Plots similar to Fig.\ \ref{fig_3lyr}
for ABA fivelayer graphene.
In (a), the dotted curve indicates energy bands calculated
by neglecting the off-diagonal block.
}
\label{fig_5lyr}
\end{figure}

Fig.\ \ref{fig_5lyr} (a) and (b) show
the zero-field band structure, and
the  Landau levels of fivelayer graphene, respectively
numerically computed with all the parameters.
Here the symbols `M', `b' and `B'
represent the monolayer $(m=0)$, the light-mass bilayer
$(m=2)$ and the heavy-mass bilayer block $(m=4)$, respectively.
Fig.\ \ref{fig_5lyr}(c) illustrates a
two-dimensional map of the density of states
and the quantum Hall conductivity.
Similarly to the trilayer,
the quantized Hall conductivity jumps by $2$
only at the valley-split
zero energy levels of monolayer-like and bilayer-like subbands,
while it changes in units of 4 at other levels
which are almost valley degenerate.
As a result, the quantized Hall conductivity
becomes $4M$ only in the region between
monolayer zero-levels of two valleys,
and in the narrow gaps between
$-1$st and $0$th levels of each bilayer band.
Otherwise it takes a series of $4M+2$.

\section{Conclusion}

Here, we used the effective mass approximation to analyze
the Landau level spectra of ABA-stacked multilayer graphenes.
In general, the next-nearest layer couplings
significantly influence the low-energy spectrum and the
quantum Hall effect by causing energy shifts, level anti-crossings,
and valley splitting of the low-lying Landau levels,
as described in detail here for trilayer, fourlayer and fivelayer graphene.
We considered a single-particle picture in order to provide a simple
description of a broad range of features. Depending on sample quality, electron-electron
interactions will also contribute to symmetry breaking and Landau level splitting
\cite{zhang06,giesbers09,feldman09,zhao09,du09,dean10a,dean10b,min08b,zhang10,vafek10,nand10a,nand10b,lemonik10},
as will interlayer asymmetry due to the presence of an external gate or
doping \cite{mcc06a,mcc06b,castro07,muchassc,kosh09a,kosh10a,fogler10}.
Nevertheless, an experimental observation of features
related to next-nearest layer couplings should be possible in
high mobility samples including
suspended graphene \cite{bolotin08,du08,bol08,feldman09,lau10}
or graphene on a boron nitride substrate \cite{dean10a,dean10b,young10,pablo}.
In fact, Landau level crossings in trilayer graphene were recently
observed \cite{pablo}, and they allowed the determination of
Slonczewski-Weiss-McClure parameter values
in remarkably close agreement with those of bulk graphite \cite{trivalues,dressel02}.

\section{Acknowledgments}

The authors thank T.~Taychatanapat and P.~Jarillo-Herrero
for discussions and for sharing their experimental data prior to publication.
This project has been funded by JST-EPSRC Japan-UK Cooperative Programme Grant EP/H025804/1.


\begin{thebibliography}{99}
\bibitem{kpoints} Corners of the hexagonal Brillouin zone are located at
wave vector $\mathbf{K}_{\xi }=\xi ({\textstyle\frac{4}{3}}\pi a^{-1},0)$,
where $\xi =\pm 1$ and $a$ is the lattice constant.
\bibitem{trivalues}
Recent experiments \cite{pablo} on high-mobility trilayer samples
determined Slonczewski-Weiss-McClure parameter values to be
$\gamma_2 = -0.028$~eV, $\gamma_4 = 0.041$~eV,
$\gamma_5 = 0.05$~eV and $\Delta' = 0.046$~eV, 
when $\gamma_0 = 3.1$~eV, $\gamma_1 = 0.39$~eV and $\gamma_3 = 0.315$~eV
are assumed.
\end{thebibliography}
\end{document}